\documentclass[conference]{IEEEtran}
\IEEEoverridecommandlockouts

\usepackage[cmex10]{amsmath}
\usepackage[utf8x]{inputenc}
\usepackage[T1]{fontenc}
\usepackage[american]{babel}
\usepackage{textcomp, amsmath, amssymb, amsfonts, amsthm, bm, bbm, ucs, cite, booktabs, multirow, algorithmic, algorithm, graphicx, siunitx, balance}
\usepackage[table]{xcolor}

\usepackage{fancyhdr}
\IEEEoverridecommandlockouts
\fancypagestyle{myfancy}{
	\fancyhf{} 
	\fancyhead[C]{\textit{Accepted to the 33rd European Signal Processing Conference (EUSIPCO 2025), Isola delle Femmine, Palermo, Italy}} 
	\fancyfoot[C]{\footnotesize © 2025 IEEE. Personal use of this material is permitted. 
	Permission from IEEE must be obtained for all other uses.}

}

\graphicspath{{./figures/},{./imgs/}}

\theoremstyle{definition}

\theoremstyle{remark}

\newcommand*{\herm}{^{\mathsf{H}}}
\newcommand*{\transp}{^{\mathsf{T}}}

\newcommand{\e}{\mathrm{e}}
\renewcommand{\i}{\mathrm{i}}

\DeclareMathOperator*{\argmax}{\arg\,\max}

\allowdisplaybreaks
\definecolor{lightpink}{rgb}{1, 0.88, 0.88}

\begin{document}
\bstctlcite{BSTcontrol}	

\title{STAR-RIS Transceivers: Integrated Sensing and Communication with Pulsed Signals}

\author{
\IEEEauthorblockN{Hedieh~Taremizadeh\IEEEauthorrefmark{1}\IEEEauthorrefmark{2}, 
Emanuele Grossi\IEEEauthorrefmark{1}\IEEEauthorrefmark{2}\IEEEauthorrefmark{3},  Luca  Venturino\IEEEauthorrefmark{1}\IEEEauthorrefmark{2}\IEEEauthorrefmark{3}}
\IEEEauthorblockA{\IEEEauthorrefmark{1}\textit{University of Cassino and Southern Lazio, 03043 Cassino, Italy}}
\IEEEauthorblockA{\IEEEauthorrefmark{2}\textit{European University of Technology EUt+, European Union}}
\IEEEauthorblockA{\IEEEauthorrefmark{3}\textit{National Inter-University Consortium for Telecommunications, 43124 Parma, Italy}}
\IEEEauthorblockA{E-mails: hedieh.taremizadeh@unicas.it,  e.grossi@unicas.it, l.venturino@unicas.it. }
\thanks{The work of H.~Taremizadeh was supported by the Italian Ministry of Education, University, and Research with the program ``Dipartimenti di Eccellenza 2018--2022''. The work of E. Grossi was supported by the project ``CommunIcations and Radar Co-Existence (CIRCE),'' CUP H53D23000420006, funded by the European Union under the Italian National Recovery and Resilience Plan of NextGenerationEU (PE00000001, program ``RESTART'').  The work of L. Venturino was supported by the European Union under the Italian National Recovery and Resilience Plan (PNRR) of NextGenerationEU partnership on ``Telecommunications of the Future'' (PE00000001 -- program ``RESTART''), CUP E63C22002040007 -- D.D. n.~1549 of 11/10/2022.}
}
\maketitle

\thispagestyle{myfancy}

\begin{abstract}
This study examines an integrated sensing and communication (ISAC) transceiver featuring a simultaneous transmitting and reflecting reconfigurable intelligent surface (STAR-RIS) and a receiver equipped with a passive electronically scanned array (PESA) and a single digital channel. By utilizing a periodic pulsed signal emitted by a feeder, we introduce at the STAR-RIS a space modulation to illuminate two angular directions observed by the radar receiver, one in each half-space, and a time modulation to distinguish the corresponding echoes from prospective moving targets and embed communication messages. The proposed time modulation employs orthogonal binary codebooks with different trade-offs in transmission and error rates, while having minimal impact on the radar performance, evaluated by probability of detection and root mean square error in the radial velocity estimation.
\end{abstract}

\section{Introduction}\label{SEC_Introduction}
A reconfigurable intelligent surface (RIS) is a planar array of sub-wavelength elements that can be reconfigured to manipulate electromagnetic waves cost-effectively. RISs have gained significant attention for beyond-5G wireless communications, as they can be used to enhance security, reliability, and both spectral and energy efficiency~\cite{Bjornson-2022, Ye_2022}, possibly with the aid of artificial intelligence~\cite{Wang_2021}. They can also be employed to implement backscatter communications~\cite{Liang_2022, Venturino-2023}, acting either as a backscatter device (tag) or as a helper to improve the source-tag-destination channel, to improve target detection and localization in radar applications~\cite{Taremizadeh-2023, Lops-2024}, and, more generally, to provide better system trad-offs in integrated sensing and communication (ISAC)~\cite{Grossi-arXiv2024, Magbool-arXiv2024}. Traditional RISs operate in either a reflective-only or transmissive-only mode, limiting their coverage to a single half-space. To overcome this limitation, simultaneous transmitting and reflecting RISs (STAR-RISs) have also been introduced~\cite{Manzoor-2023, Umer-2025}. The hardware design of STAR-RISs is analyzed in~\cite{Xu_2021}, while some practical operational modes, namely, energy splitting, mode switching, and time switching, are discussed in~\cite{Schober-STAR-RIS-2022}. 

\begin{figure}[t]
\centering
\includegraphics[width=0.8\columnwidth]{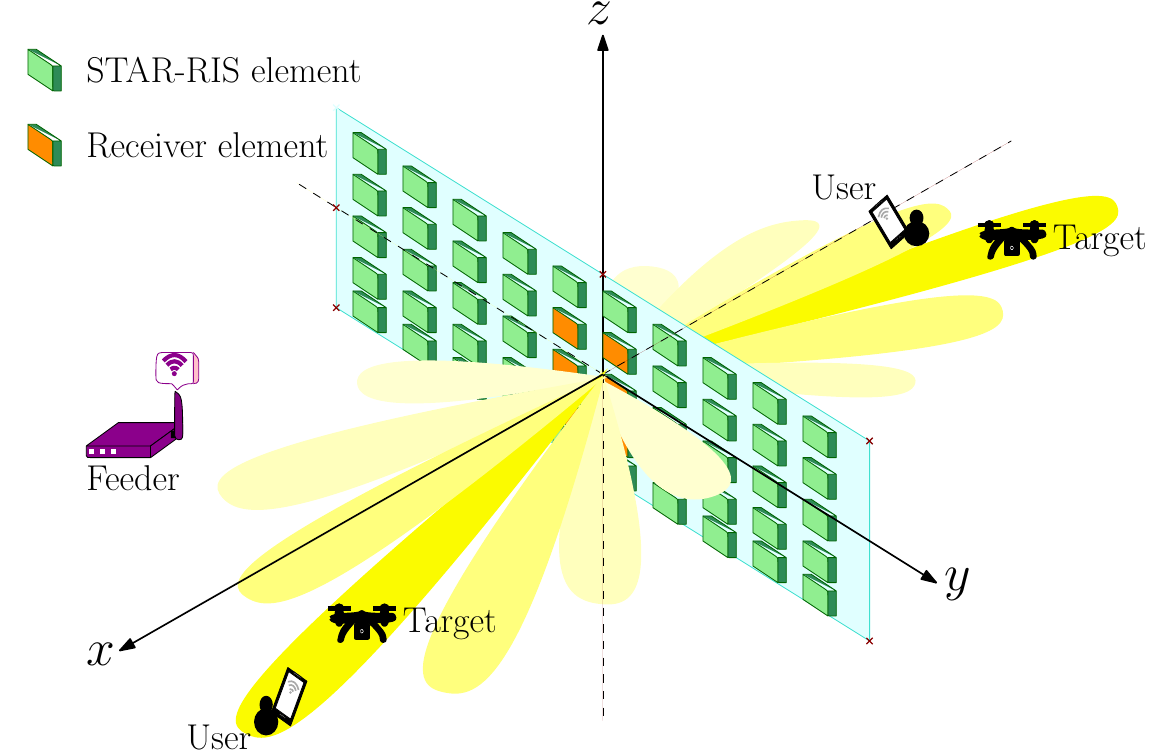}
\caption{System architecture.}
\label{Fig_01}
\vspace*{-0.25cm}
\end{figure}

In this study, we consider the ISAC transceiver architecture in Fig.~\ref{Fig_01}, encompassing a low-cost feeder emitting a periodic pulsed signal, a STAR-RIS acting as a dual-function transmitter, and a radar receiver collocated with the STAR-RIS. This architecture resembles the STAR-RIS-based pulsed Doppler radar considered in~\cite{Taremizadeh-2024-EUSIPCO}, but it differs in several respects: first, the communication function is now integrated at the STAR-RIS; second, we assume that the radar receiver is equipped with a passive electronically scanned array (PESA) and a single digital channel, rather than a fully digital array, to reduce implementation costs. The main contribution is the introduction of a space-time coded modulation at the STAR-RIS that aims to illuminate two distinct angular directions monitored by the radar receiver, with one direction in each half-space, while also embedding communication messages within the redirected signals. A key design challenge is that no channel state information (CSI) is assumed at the user side; also, the echoes from moving targets located in different half-spaces must be distinguishable by the radar receiver that implements a decision rule based on the generalized-information criterion (GIC)~\cite{Stoica-2004}. The proposed time modulation leverages orthogonal binary codebooks and has a minimal impact on the radar's performance, which is evaluated based on the probability of detection and the root mean square error in estimating the radial velocity of targets, making it a promising solution for future radar-centric ISAC applications.

The remainder of this paper is organized as follows. Section~\ref{SEC_Description} describes the system model.\footnote{Vectors and matrices are denoted by bold lowercase and uppercase letters, respectively. The symbols $\i$, $(\cdot)^*$, $(\cdot)\transp$, $(\cdot)\herm$ represent, the imaginary unit, conjugate, transpose, conjugate-transpose, respectively.  $\bm{I}_{M}$ is the $M\times M$ identity matrix. $\mathbbm 1_{\mathcal A}$ is the indicator function of the condition $\mathcal A$ and $\text{diag}\{\bm{x}\}$ represents a diagonal matrix with $\bm{x}$ on its main diagonal.
The symbols $\lceil \cdot \rceil$ and $\ast$ denote the ceiling function  and the convolution operation, respectively.}
Section~\ref{SEC_System_Design} discusses the STAR-RIS response design and the radar and communication receivers. Section~\ref{SEC_Results} presents the numerical analysis. Finally, the conclusions are drawn in Section~\ref{SEC_Conclusions}.

\section{System description}\label{SEC_Description}
Consider the system in Fig.~\ref{Fig_01} encompassing a feeder operating with a carrier frequency $f_{o}$, a planar STAR-RIS composed of $N_{\rm ris}$ elements, a radar receiver collocated with the STAR-RIS and equipped with a PESA composed of $N_{\rm rad}$ quasi-omnidirectional antennas, and a single-antenna user on each side of the STAR-RIS.  The feeder emits the baseband pulse train $\sum_{p=-\infty}^{+\infty}\sqrt{\mathcal{P}\Delta}\psi(t - pT + \delta)$, where $\mathcal{P}$ is the pulse power, $T$ is the pulse repetition interval (PRI), $\delta$ is time of flight between the feeder and STAR-RIS, and $\psi(t)$ is a unit-energy pulse with bandwidth $B$ that is zero if $t\notin[0, \Delta)$, with $\Delta \approx 1/B$ and $\Delta \ll T$. After superimposing a space-time modulation, the STAR-RIS redirects the signal from the feeder towards both the transmissive and reflective half-spaces: the goal is to illuminate the angular directions monitored by the radar receiver while sending a message to the users. The PESA at the radar receiver simultaneously collects the echoes from both half-spaces and, after analog beamforming, processes the signal via a single digital channel over a coherent processing interval (CPI) spanning $P$ PRIs; without loss of generality, the CPI $[0, PT]$ is considered next for illustration.

\subsection{Design assumptions}

We consider uniform square arrays with half-wavelength spacing,  neglecting any coupling among their elements. A narrowband assumption is made, and far-field propagation is considered. The channel between the feeder and STAR-RIS is known to the latter and constant over a CPI. Finally, we assume perfect synchronization between the feeder and radar receiver, but no CSI at the users.  

For future reference, we denote by $\bm{g} \in \mathbb{C}^{N_{\rm ris}}$ the channel between the feeder and STAR-RIS. In the Cartesian reference system in Fig.~\ref{Fig_01}, we denote by  $\bm{\phi}=[\phi^{\rm az};\phi^{\rm el}]$ the angular direction with azimuth angle $\phi^{\rm az}$ and elevation angle $\phi^{\rm el}$; accordingly, $\mathcal{F}_{\rm tr} = \left(\frac{\pi}{2}, \frac{3\pi}{2}\right) \times \left(-\frac{\pi}{2}, \frac{\pi}{2}\right)$ and  $\mathcal{F}_{\rm re} = \left(-\frac{\pi}{2}, \frac{\pi}{2}\right) \times \left(-\frac{\pi}{2}, \frac{\pi}{2}\right)$ are the sets of angular directions in the transmissive and reflective half-spaces, respectively. Finally, we denote by $G_{\rm ris}(\bm{\phi})$ and $G_{\rm rad}(\bm{\phi})$ the element gains of the STAR-RIS and radar receiver  towards $\bm{\phi}$, respectively, and by  $\bm{u}_{\rm ris}(\bm{\phi})\in \mathbb{C}^{N_{\rm ris}}$ and $\bm{u}_{\rm rad}(\bm{\phi})\in \mathbb{C}^{N_{\rm rad}}$  the steering vectors of the STAR-RIS and radar receiver towards $\bm{\phi}$, respectively.

\subsection{STAR-RIS responce}\label{sub_SEC_Response}

In the $p$-th PRI, for $p=0,\ldots,P-1$, the STAR-RIS response in the transmissive and reflective half-spaces is modeled as $\bm{x}_{\rm tr}(p)= c_{\rm tr}(p) \bm{s}_{\rm tr}$ and $\bm{x}_{\rm re}(p)= c_{\rm re}(p) \bm{s}_{\rm re}$, respectively, where $\bm{s}_{\rm tr}$ and $\bm{s}_{\rm tr}$ are  $N_{\rm ris}$--dimensional spatial beamformers with unit-magnitude entries and $\bm{c}_{\rm tr}=[c_{\rm tr}(0);\cdots;c_{\rm tr}(P-1)]$ and $\bm{c}_{\rm re}=[c_{\rm re}(0);\cdots;c_{\rm re}(P-1)]$  are $P$--dimensional code sequences with $|c_{\rm tr}(p)|^2 + |c_{\rm re}(p)|^2 = 1$. Hence, the signal redirected by the STAR-RIS towards an angular direction $\bm{\phi}$ in the CPI under inspection is
\begin{equation}
\zeta_{\rm RIS} (t, \bm{\phi}) = \zeta_{\rm tr} (t, \bm{\phi}) + \zeta_{\rm re} (t, \bm{\phi}), \label{redirected-signal}
\end{equation}
for $t \in [0,PT]$, where  
\begin{align}
\zeta_{\rm i} (t, \bm{\phi}) &= \mathbbm{1}_{\{\bm{\phi} \in \mathcal{F}_{\rm i}\}} \sqrt{\mathcal{P}\Delta G_{\rm ris}(\bm{\phi})} \Bigl(\bm{u}_{\rm ris}^{T}(\bm{\phi}) \mathrm{diag}(\bm{g}) \bm{s}_{\rm i} \Bigr) \notag \\
 &\quad \times \sum_{p=0}^{P-1}  c_{\rm i}(p) \psi(t - pT),\quad \rm i\in \{\rm tr, re \}.
\end{align}%
The design of the space-time STAR-RIS response over a CPI will be discussed in Sec.~\ref{SEC_System_Design}.

\subsection{Radar received signal}\label{sub_SEC_Rad_received_signal}
The radar receiver monitors the angular directions $\bm{\phi}_{\rm tr,t}=[\phi_{\rm tr,t}^{\rm az};\phi_{\rm tr,t}^{\rm el}]\in \mathcal{F}_{\rm tr}$ and $\bm{\phi}_{\rm re,t}=[\phi_{\rm re,t}^{\rm az,t};\phi_{\rm re,t}^{\rm el}]\in \mathcal{F}_{\rm re}$. These directions are chosen to be mirror-symmetric relative to the plane containing the STAR-RIS and radar receiving array: specifically, $\phi_{\rm re,t}^{\rm az}=\pi-\phi_{\rm tr,t}^{\rm az}$ and $\phi_{\rm re,t}^{\rm el}=\phi_{\rm tr,t}^{\rm el}$. This choice is motivated by the fact that, for a planar uniform PESA, $\bm{u}_{\rm rad}([\phi^{\rm az};\phi^{\rm el}])=\bm{u}_{\rm rad}([\pi-\phi^{\rm az};\phi^{\rm el}])$ for any $[\phi^{\rm az};\phi^{\rm el}]\in\mathcal{F}_{\rm re}$.

After pulse compression and range gating, the received signal at the range of interest can be written as~\cite{Taremizadeh-2024-EUSIPCO}
\begin{align}&y_{\rm rad}(p)=z_{\rm rad}(p)\notag\\
	&\quad+\alpha_{\rm tr,t}  \sqrt{\frac{\mathcal{P}\Delta G_{\rm rad}(\bm{\phi}_{\rm tr,t}) G_{\rm ris}(\bm{\phi}_{\rm tr,t}){\lambda^2}}{4 \pi}} \bigl(\bm{s}_{\rm rad}\herm\bm{u}_{\rm rad}(\bm{\phi}_{\rm tr,t})\bigr) \notag \\ 
	&\quad \times\bigl(\bm{u}_{\rm ris}\transp(\bm{\phi}_{\rm tr,t})\mathrm{diag}\{\bm{g}\}\bm{s}_{\rm tr}\bigr)c_{\rm tr}(p)\e^{\i 2\pi \nu_{\rm tr,t}T (p-1)} \notag \\
	&\quad+\alpha_{\rm re,t} \sqrt{\frac{\mathcal{P}\Delta G_{\rm rad}(\bm{\phi}_{\rm re,t}) G_{\rm ris}(\bm{\phi}_{\rm re,t}){\lambda^2}}{ 4 \pi}} \bigl(\bm{s}_{\rm rad}\herm\bm{u}_{\rm rad}(\bm{\phi}_{\rm re,t})\bigr) \notag \\ 
	&\quad \times \bigl(\bm{u}_{\rm ris}\transp(\bm{\phi}_{\rm re,t})\mathrm{diag}\{\bm{g}\}\bm{s}_{\rm re}\bigr)c_{\rm re}(p)\e^{\i 2\pi \nu_{\rm re,t}T (p-1)},\label{eq-observation-1}
\end{align}
for $p=0,\ldots, P-1$, where: $z_{\rm rad}(p)$ is the disturbance; $\bm{s}_{\rm rad}\in\mathbb{C}^N_{\rm rad}$ is the analog beamformer of the PESA with unit norm and constant modulus entries; $\alpha_{\rm tr,t}\in\mathbb{C}$ and $\nu_{\rm tr,t}\in\mathbb{R}$ are the unknown amplitude and Doppler shift of a prospective target in the transmissive side, respectively; $\alpha_{\rm re,t}\in\mathbb{C}$ and $\nu_{\rm re,t}\in\mathbb{R}$ are the unknown amplitude and Doppler shift of a prospective target in the reflective side, respectively. Here, $\alpha_{\rm tr,t}$  and $\alpha_{\rm re,t}$ account for the two-way path-loss from the STAR-RIS to the target and the target radar cross-section, and, for $\alpha_{\rm tr,t}=0$ and/or $\alpha_{\rm re,t}=0$  no target is present in the transmissive and/or reflective side, respectively. 

The samples in~\eqref{eq-observation-1} are organized into the $P$--dimensional vector $\bm{y}_{\rm rad}=\big[y_{\rm rad}(0);\cdots;y_{\rm rad}(P-1)\big]$. For any $\bm{c}=[c(0);\cdots;c(P-1)] \in \mathbb{C}^{P}$ and $\nu \in \mathbb{R}$, define
\begin{equation}
    \bm{h}(\bm{c}, \nu) = \big[c(0); c(1) e^{\i 2\pi T \nu}; \dots; c(P-1) e^{\i 2\pi \nu T (P-1)} \big];
\end{equation}  
also, for any $\bm{\phi} \in \mathcal{F}_{\rm i}$ and $\rm i\in \{\rm tr, re \}$, let 
\begin{align}\label{eq:gamma_def}
    \gamma_{\rm i}(\bm{\phi}, \bm{s}_{\rm i}, \bm{s}_{\rm rad}) &= \sqrt{\mathcal{P}\Delta G_{\rm rad}(\bm{\phi}) G_{\rm ris}(\bm{\phi})\lambda^2/(4\pi)} 
    \bigl(\bm{s}_{\rm rad}^\mathsf{H} \bm{u}_{\rm rad}(\bm{\phi})\bigr)\notag   \\
    &\quad \times \bigl(\bm{u}_{\rm ris}^\mathsf{T}(\bm{\phi}) \operatorname{diag}(\bm{g}) \bm{s}_{\rm i}\bigr).
\end{align}
Then, we can write
\begin{align}
    \bm{y}_{\rm rad} &= \alpha_{\rm tr,t} \gamma_{\rm tr}(\bm{\phi}_{\rm tr,t}, \bm{s}_{\rm tr}, \bm{s}_{\rm rad}) \bm{h}(\bm{c}_{\rm tr}, \nu_{\rm tr,t}) \notag\\
    &\quad + \alpha_{\rm re,t} \gamma_{\rm re}(\bm{\phi}_{\rm re,t}, \bm{s}_{\rm re}, \bm{s}_{\rm rad}) \bm{h}(\bm{c}_{\rm re}, \nu_{\rm re,t}) + \bm{z}_{\rm rad},
    \label{eq:y_radar}
\end{align}  
where $\bm{z}_{\rm rad} = \big[z_{\rm rad}(0);\cdots;z_{\rm rad}(P-1)\big]$.  As in~\cite{Taremizadeh-2024-EUSIPCO}, $\bm{z}_{\rm rad}$ is modeled as a circularly-symmetric complex Gaussian vector with known full-rank covariance matrix $\bm{C}_{\rm rad}$.

\subsection{Received signal at the users}\label{SEC_rx_signal_user}
When the communication function is present, the STAR-RIS partitions the CPI into $P/M$ time slots of equal duration, where $M$ is chosen so that $P/M$ is an integer and $MT$ is smaller than the coherence time of the channel from the STAR-RIS to the user. In every time slot, the STAR-RIS broadcasts a message to each half-space (more on this in Sec.~\ref{SEC_System_Design}). 

For illustration, consider the time slot $[0,MT]$ and the transmissive half-space. Then, the signal at the user is
\begin{equation}\label{user-rx-signals}
y_{\rm tr,u}(t) = \sum_{k=1}^{K_{\rm tr,u}} \alpha_{{\rm tr,u},k} \zeta_{\rm tr} (t-\tau_{{\rm tr,u},k}, \bm{\phi}_{{\rm tr,u},k}) + z_{\rm tr, u}(t),
\end{equation}
for $t\in[0,MT]$, where: $K_{\rm tr,u}\geq 1$ is the number of paths; $\alpha_{{\rm tr,u},k}\in \mathbb{C}$ is the amplitude of the $k$-th path; $\tau_{{\rm tr,u},k}\in[\tau_{\rm tr,u,min}, \tau_{\rm tr,u,max}]$ is the delay of the $k$-th path, with $0\leq \tau_{\rm tr,u,min}\leq\tau_{\rm tr,u,max}\leq T-2\Delta$; $\bm{\phi}_{{\rm tr,u},k}\in\mathcal F_{\rm tr}$ is the angular direction of departure of the $k$-th path; $w_{\rm tr, u}(t)$ is the noise. 

The signal in~\eqref{user-rx-signals} is passed through a unit-energy filter matched to $\psi(t)$ and sampled at the Nyquist rate $1/B$. Upon denoting by  $r_{\psi}(t)=\psi(t)\ast \psi^*(-t)$ the autocorrelation function of $\psi(t)$, the output samples are
\begin{align}
y_{{\rm tr},u}(p, \ell)  &= y_{\rm tr,u}(t) \ast \psi^*(-t) \mid_{t=\tau_{\rm tr,u,min}+pT+\ell/B} \notag\\
&=c_{\rm tr}(p) \beta_{{\rm tr},u}(\ell)  + z_{{\rm tr},u}(\ell, p), \label{rx_signal_user}
\end{align}
for $p=0,\ldots,M-1$ and $\ell=0,\ldots,L_{\rm tr, u}-1$, where $L_{\rm tr, u}=\lceil (\tau_{\rm tr,u,max}-\tau_{\rm tr,u,min}+2\Delta)B\rceil$ and
\begin{subequations}
\begin{align}
\beta_{{\rm tr},u}(\ell)  &=\sum_{k=1}^{K_{\rm tr,u}} \alpha_{{\rm tr,u},k} \sqrt{\mathcal{P}\Delta G_{\rm ris}(\bm{\phi}_{{\rm tr,u},k})} \notag \\
&\quad \times r_{\psi}\big(\tau_{\rm tr,u,min}
+\ell/B-\tau_{{\rm tr,u},k}\big)\notag \\
&\quad \times \bigl(\bm{u}_{\rm ris}\transp(\bm{\phi}_{{\rm tr,u},k})\mathrm{diag}\{\bm{g}\}\bm{s}_{\rm tr}\bigr), \label{rx_signal_user_beta} \\
z_{{\rm tr},u}(p, \ell) &= z_{\rm tr,u}(t) \ast \psi^*(-t) \mid_{t=\tau_{\rm tr,u,min}+pT +\ell/B}.
\end{align}
\end{subequations}
The coefficients $\{\beta_{{\rm tr},u}(\ell)\}_{\ell=0}^{L_{\rm tr, u}-1}$ are the tap amplitudes of the overall discrete-time channel from the feeder to the STAR-RIS to the user. Also, since $r_{\psi}(t)=0$ if $|t|\geq\Delta$, only the terms for which $|\tau_{\rm tr,u,min} +\ell/B-\tau_{{\rm tr,u},k}|<\Delta$ contribute to~\eqref{rx_signal_user_beta}.

The samples in~\eqref{rx_signal_user} are organized into the matrix $\bm{Y}_{{\rm tr},u}\in\mathbb{C}^{M\times L_{\rm tr, u}}$, with $[\bm{Y}_{{\rm tr},u}]_{p+1,\ell+1}=y_{{\rm tr},u}(p, \ell) $ for $p=0,\ldots,M-1$ and $\ell=0,\ldots,L_{\rm tr, u}-1$.  Each row (column) of $\bm{Y}_{{\rm tr},u}$ contains samples spaced $1/B$ ($T$) apart in time;  in particular, we have 
\begin{equation} \label{rx_signal_user_matrix}
\bm{Y}_{{\rm tr},u}=\bm{c}_{\rm tr}(p) \bm{\beta}_{{\rm tr},u}\transp + \bm{Z}_{{\rm tr},u},
\end{equation}
where $\bm{\beta}_{{\rm tr},u}=[\beta_{{\rm tr},u}(0);\cdots; \beta_{{\rm tr},u}(L_{\rm tr, u}-1) ]\in\mathbb{C}^{L_{\rm tr, u}}$ and $[\bm{Z}_{{\rm tr},u}]_{p+1,\ell+1}=z_{{\rm tr},u}(p, \ell) $ for $p=0,\ldots,M-1$ and $\ell=0,\ldots,L_{\rm tr, u}-1$. The entries of $\bm{Z}_{{\rm tr},u}$ are modeled as independent Gaussian random variables with variance $\sigma^{2}_{\rm com}$.

\section{System design}\label{SEC_System_Design}
The available degrees of freedom are the spatial beamformers ($\bm{s}_{\rm tr}$ and $\bm{s}_{\rm re}$) and the temporal code sequences ($\bm{c}_{\rm tr}$ and $\bm{c}_{\rm re}$) of the STAR-RIS, the implementation of the radar receiver (including both the analog beamformer $\bm{s}_{\rm rad}$ and the decision rule), and the implementation of the communication receiver. Next, we will discuss each of these aspects.

\subsection{Design of $\bm{s}_{\rm tr}$ and $\bm{s}_{\rm re}$}
Since the radar function is the primary task, the spatial beamformer $\bm{s}_{\rm tr}$ is chosen to maximize the array power gain of the STAR-RIS towards the angular direction inspected by the radar receiver in the transmissive half-space, namely, $|\bm{u}_{\rm ris}\transp(\bm{\phi}_{\rm tr,t})\mathrm{diag}\{\bm{g}\}\bm{s}_{\rm tr}|^2$. The spatial beamformer $\bm{s}_{\rm re}$ is similarly constructed. For $n=1,\ldots,N_{\rm ris}$,  we have~\cite{Taremizadeh-2024-EUSIPCO}:  
\begin{subequations}
\begin{align}
[\bm{s}_{\rm tr}]_{n}&=\e^{-\i(\angle[\bm{g}]_{n}+\angle[\bm{u}_{\rm ris}(\bm{\phi}_{\rm tr,t})]_{n})},\\
[\bm{s}_{\rm re}]_{n}&=\e^{-\i (\angle[\bm{g}]_{n}+\angle[\bm{u}_{\rm ris}(\bm{\phi}_{\rm re,t})]_{n})}.
\end{align}
\end{subequations}

\subsection{Design of $\bm{c}_{\rm tr}$ and $\bm{c}_{\rm re}$} 
The temporal code sequences must ensure that the echoes from different half-spaces can be distinguished by the radar receiver~\cite{Taremizadeh-2024-EUSIPCO} (see Sec.~\ref{Sec:Radar_rx_design}); therefore, $\bm{c}_{\rm tr}$ and $\bm{c}_{\rm re}$ should have a low cross-correlation and be Doppler tolerant. Also, if the communication function is active, they must embed a message that can be decoded without CSI (see Sec.~\ref{Sec:User_rx_design}). This is a challenging problem whose comprehensive analysis is left for future development; next, we will discuss a heuristic solution inspired by previous encoding strategies considered in~\cite{Taremizadeh-2024-EUSIPCO} and~\cite{Venturino-2023} for radar-only and communication-only applications, respectively.

Assume that $P$ is a power of $2$. When the communication function is active, we propose to construct  $\bm{c}_{\rm tr}$ and $\bm{c}_{\rm re}$ as the concatenation of $P/M$ codewords taken from distinct codebooks $\mathcal{C}_{\rm tr}$ and $\mathcal{C}_{\rm re}$, respectively, so that $[{\bm c}_{\rm tr}]_{(m-1)M+1:mM}\in \mathcal{C}_{\rm tr}$ and $[{\bm c}_{\rm re}]_{(m-1)M+1:mM}\in \mathcal{C}_{\rm re}$, for $m=1,\ldots,P/M$. Let $M\geq 2^{b+1}$, where $b> 0$ is the number of bits sent in each half-space during a time slot; also, let $\bm H_n \in\{-1,1\}^{n\times n}$ be the Hadamard of order $n$. Then we propose to have\footnote{If $P$ is a power of $2$ and $P/M$ integer, then $M$ is also a power of $2$ and therefore $\bm H_{M}$ exists.}
\begin{subequations}\label{codebooks}
 \begin{align}
  \mathcal{C}_{\rm tr}&=\bigl\{\text{first } 2^b \text{ columns of } (1/\sqrt{2})\bm H_{M} \bigr\},\\
  \mathcal{C}_{\rm re}&=\bigl\{ \text{last } 2^b \text{ columns of }(1/\sqrt{2})\bm H_{M} \bigr\}.
 \end{align}%
\end{subequations} Notice that each codebook contains orthogonal codewords, thus facilitating the message recovery. Also,  the factor $1/\sqrt{2}$ in~\eqref{codebooks} is included to ensure energy conservation at the STAR-RIS. Finally, we have observed that using the columns of the Hadamard matrix from right to left in $\mathcal{C}_{\rm tr}$ and from left-to-right in $\mathcal{C}_{\rm re}$ guarantees Doppler tolerance.

When the communication function is not active, we choose $\bm{c}_{\rm tr}$ and $\bm{c}_{\rm re}$ as the first and the last column of $\bm H_P$, respectively, which corresponds to set $M=P$ and $b=0$ in~\eqref{codebooks}. 

\subsection{Implementation of the radar receiver}\label{Sec:Radar_rx_design}
The analog beamformer $\bm{s}_{\rm rad}$ is chosen to maximize the array power gain of the PESA towards the inspected angular directions, namely, $|\bm{s}_{\rm rad}\herm\bm{u}_{\rm rad}(\bm{\phi}_{\rm tr,t})|^2=|\bm{s}_{\rm rad}\herm\bm{u}_{\rm rad}(\bm{\phi}_{\rm re,t})|^2$, where the latter equality is a consequence of the angular directions being mirror-symmetric relative to the array plane. Hence, 
we have $\bm{s}_{\rm rad}\propto \bm{u}_{\rm rad}(\bm{\phi}_{\rm tr,t})$.

Then, the radar detector exploits  $\bm{c}_{\rm tr}$ and $\bm{c}_{\rm re}$ to distinguish and detect prospective echoes from the two half-spaces. With reference~\eqref{eq:y_radar}, there are four hypotheses:
\begin{description}
    \item[\hspace{8pt}$\mathcal{H}_0$:] No target is present on both sides;
    \item[$\mathcal{H}_{1, \rm tr}$:] The target is only present on the transmissive side;
    \item[$\mathcal{H}_{1, \rm re}$:] The target is only present on the reflective side;
    \item[\hspace{9pt}$\mathcal{H}_2$:] The target is present on both sides.
\end{description}
As in~\cite{Taremizadeh-2024-EUSIPCO}, we resort to a GIC-based decision rule~\cite{Stoica-2004}. To proceed, define the following quantities:
\begin{subequations}
\begin{align}
\bm{\xi}_{\rm tr}(\nu_{\rm tr,t})&=\bm{C}_{\rm rad}^{-1}\bm{h}(\bm{c}_{\rm tr}, \nu_{\rm tr,t})/\|\bm{C}_{\rm rad}^{-\frac{1}{2}}\bm{h}(\bm{c}_{\rm tr}, \nu_{\rm tr,t})\|,\\
\bm{\xi}_{\rm re}(\nu_{\rm re,t})&=\bm{C}_{\rm rad}^{-1}\bm{h}(\bm{c}_{\rm re}, \nu_{\rm re,t})/\|\bm{C}_{\rm rad}^{-\frac{1}{2}}\bm{h}(\bm{c}_{\rm re}, \nu_{\rm re,t})\|,\\
\!\!\!\bm{\Xi}(\nu_{\rm tr,t},\nu_{\rm tr,t})&=\bm{C}_{\rm rad}^{-1}\bm{H}_{\rm rad}\big(\bm{H}_{\rm rad}\herm \bm{C}_{\rm rad}^{-1}\bm{H}_{\rm rad}\big)^{-1/2},
\end{align}
\end{subequations}
where $\bm{H}_{\rm rad}= \begin{bmatrix}
 \bm{h}(\bm{c}_{\rm re}, \nu_{\rm re,t}) & 
\bm{h}(\bm{c}_{\rm tr}, \nu_{\rm tr,t})
\end{bmatrix}$.
Also, denote by $\mathcal{V}_{\rm tr,t}$ and $\mathcal{V}_{\rm re,t}$ the feasible Doppler sets for a prospective target in the transmissive and reflective half-spaces, respectively. Then, the selected hypothesis is~\cite{Taremizadeh-2024-EUSIPCO}
\begin{equation}\label{GIC-rule-2}	\hat{\mathcal{H}}=\arg\max_{\mathcal{L}\in\{\mathcal{H}_{0},\mathcal{H}_{1,{\rm tr}},\mathcal{H}_{1,{\rm re}},\mathcal{H}_{2}\}} \mu(\mathcal{L}),
\end{equation} 
where
\begin{equation}\label{GIC-rule-2-obj}
	\mu(\mathcal{L})\!=\!	\begin{cases}
	\!0, & \!\!\!\!\text{if } \mathcal{L}=\mathcal{H}_{0},\\  
		\displaystyle \!\max_{\nu_{\rm tr,t}\in\mathcal{V}_{\rm tr,t}} 
  \!\!\left| \bm{\xi}_{\rm tr}\herm(\nu_{\rm tr,t}) \bm{y}_{\rm rad}\right|^2
		-\eta_{\rm rad}, & \!\!\!\!\text{if } \mathcal{L}=\mathcal{H}_{1,{\rm tr}},\\
	\displaystyle \!\max_{\nu_{\rm re,t}\in\mathcal{V}_{\rm re,t}} 
  \!\!\left| \bm{\xi}_{\rm re}\herm(\nu_{\rm re,t}) \bm{y}_{\rm rad}\right|^2
		-\eta_{\rm rad}, & \!\!\!\!\text{if } \mathcal{L}=\mathcal{H}_{1,{\rm re}},\\
\displaystyle \! \max_{ \substack{\nu_{\rm tr,t}\in\mathcal{V}_{\rm tr,t} \\ \nu_{\rm re,t}\in\mathcal{V}_{\rm re,t}}} \! \!\left\| \bm{\Xi}\herm(\nu_{\rm tr,t},\nu_{\rm tr,t}) \bm{y}_{\rm rad}\right\|^2\!\!-2\eta_{\rm rad}, & \!\!\!\!\text{if } \mathcal{L}=\mathcal{H}_{2},\\
	\end{cases}
\end{equation}
and $\eta_{\rm rad}$ is a penalty factor that can be set to control the false alarm rate under $\mathcal{H}_{0}$. When a target is detected, an estimate of its Doppler shift (and hence radial velocity) is provided by the maximizer of the objective function in~\eqref{GIC-rule-2-obj} under $\hat{\mathcal{H}}$.

\subsection{Implementation of the communication receiver}\label{Sec:User_rx_design}
For illustration, consider the user in the transmissive half-space. Upon treating  $\bm{\beta}_{{\rm tr},u}$ in~\eqref{rx_signal_user_matrix} as an unknown deterministic parameter, the maximum likelihood (ML) decoding rule   selects the codeword best-correlated with the received signal~\cite{Venturino-2023}, i.e.,
\begin{equation}
\hat{\bm c}_{\rm tr}=\argmax_{\bm c \in {\mathcal C}_{\rm tr}}  \frac{\Vert\bm c\herm \bm{Y}_{{\rm tr},u} \Vert^2}{\Vert \bm c \Vert^2}. \label{estimated_codeword_tr_1}
\end{equation}

\section{Numerical Results}\label{SEC_Results}
We assume that the feeder employs rectangular pulses and consider the following system parameters: $f_{o}=28$~GHz, $B=50$~MHz,  $\mathcal{P}=30$~dBm, $T=0.25$~ms, $N_{\rm ris} = N_{\rm rad} = 256$, and $G_{\rm ris}(\bm{\phi})=G_{\rm rad}(\bm{\phi})=\frac{\pi}{4}\cos^2{\phi^{\rm az}} \cos^2{\phi^{\rm el}}$. The channel between the feeder and STAR-RIS is modeled as
\begin{equation} \label{channel_g} \bm{g} = \sqrt{ G_{\rm f} G_{\rm ris}(\bm{\phi}_{\rm f})}\frac{\lambda}{4\pi d_{\rm f}}\bm{u}_{\rm ris}(\bm{\phi}_{\rm f}), \end{equation}
where $\lambda$ is the carrier wavelength, $G_{\rm f}= 20$~dB is the antenna gain of the feeder towards the STAR-RIS, and $d_{\rm f}=3$~m and  $\bm{\phi}_{\rm f} = [-45^\circ; 0^\circ]$ are the distance and direction of the feeder from the STAR-RIS, respectively. At the radar receiver, the unambiguous Doppler interval is $\big(\!- 1/(2T), 1/(2T)\big)$, where $1/(2T)$ corresponds to a radial velocity of about $10.7$~m/s; also, the Doppler resolution is $1/(PT)$, corresponding to a radial velocity of about $1.3$~m/s for $P=16$\cite{book-Richards}; finally, we assume $\bm{C}_{\rm rad}=\sigma^{2}_{\rm rad}\bm{I}_{P}$, with $\sigma^{2}_{\rm rad}=-164$~dBm/Hz, while $\eta_{\rm rad}$ is set to have an average number of false alarms of $10^{-4}$ under $\mathcal{H}_0$. As to the targets, we assume $\bm{\phi}_{\rm tr,t} = [160^\circ; 0^\circ]$ and $\bm{\phi}_{\rm re,t} = [20^\circ; 0^\circ]$; in every snapshot, $\nu_{\rm tr,t}$ and $\nu_{\rm re,t}$ are uniformly sampled from  $\mathcal{V}_{\rm tr,t}$ and $\mathcal{V}_{\rm re,t}$, respectively, with $\mathcal{V}_{\rm tr,t}=\mathcal{V}_{\rm re,t}= (1.75\,\rm{kHz}, 2\,\rm{kHz})$; finally, we consider a Swerling I fluctuation model, whereby $\alpha_{\rm tr,t}$ and $\alpha_{\rm re,t}$ are modeled as independent circularly-symmetric Gaussian random variables with variance $\sigma^2_{\rm t}={\rm RCS}_{\rm t}/\big((4\pi)^2  d_{\rm rad}^4\big)$, where ${\rm RCS}_{\rm t}$ is the radar cross-section and $d_{\rm rad}=10$~m is the range under inspection. As to the user on the transmissive side, we assume $L_{\rm tr, u}=15$ and $K_{\rm tr, u}=3$; in every snapshot, $\{\tau_{{\rm tr,u},k}\}_{k=1}^{K_{\rm  u}}$ are uniformly sampled from  $\{\tau_{\rm tr,u,min}+\ell/B\}_{\ell=0}^{L_{\rm tr, u}-1}$, $\{\bm{\phi}_{{\rm tr,u},k}\}_{k=1}^{K_{\rm tr, u}}$ are uniformly sampled from $\mathcal{F}_{\rm tr,u}=(170^\circ, 180^\circ) \times (-25^\circ, -15^\circ)$, and $\{\alpha_{{\rm tr,u},k}\}_{k=1}^{K_{\rm tr, u}}$ are independent realizations of a circularly-symmetric Gaussian random variable with variance $\sigma^2_{\rm u}$; the same assumptions are made on the reflective side, except that the angles of departure are sampled from $\mathcal{F}_{\rm re,u}=(15^\circ, 25^\circ) \times (-25^\circ, -15^\circ)$. 

\begin{figure}[t]
\centerline{\includegraphics[width=1.1\columnwidth]{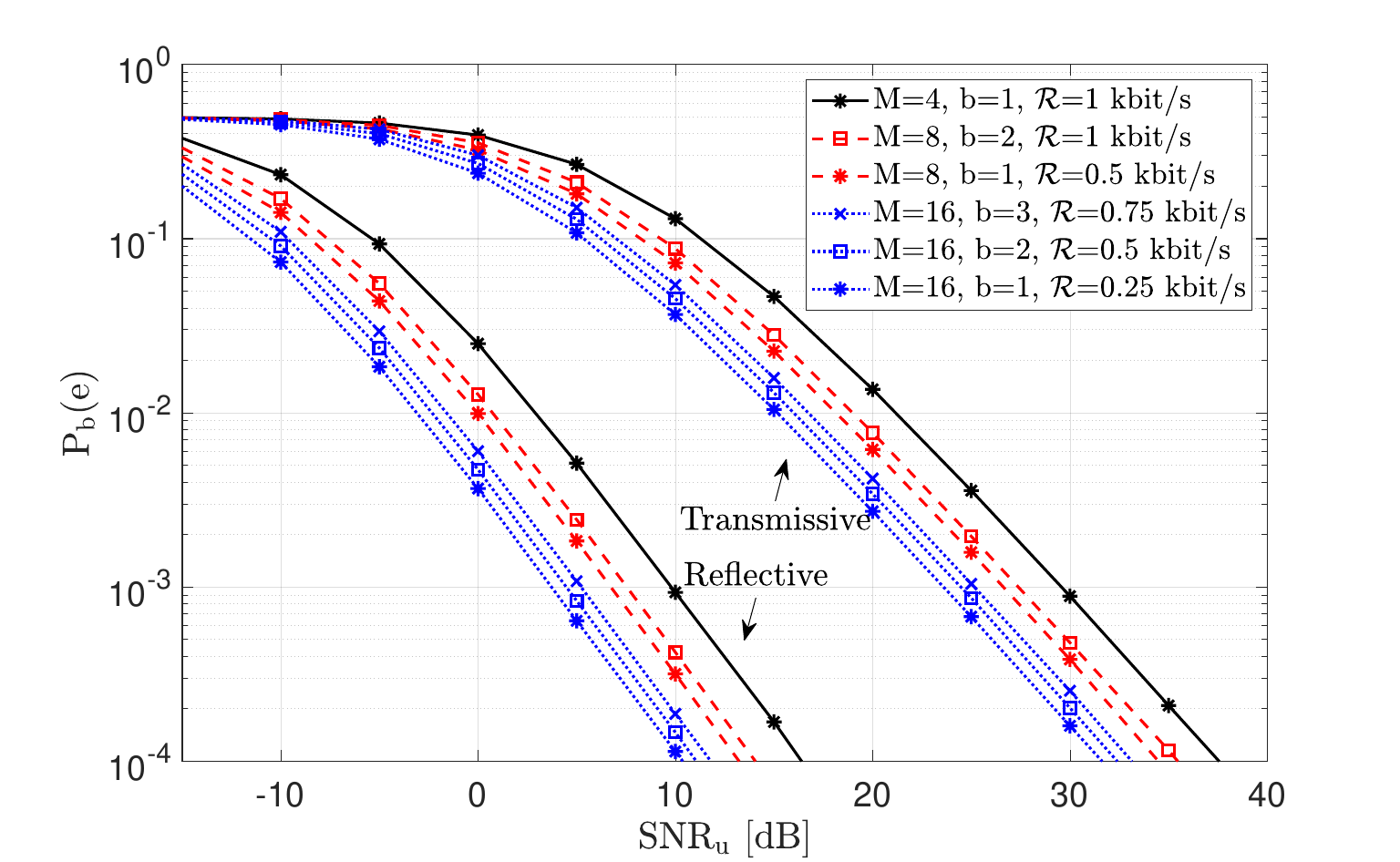}}
\vspace{-0.2cm}
\caption{BER vs $\mathrm{SNR}_{\rm u}$ for different rates and different number of pulses $M$ in the time slot.}
\vspace{-0.2cm}
\label{Fig_bit_error}
\end{figure}
\begin{figure}[t]%
\centerline{\includegraphics[width=\columnwidth]{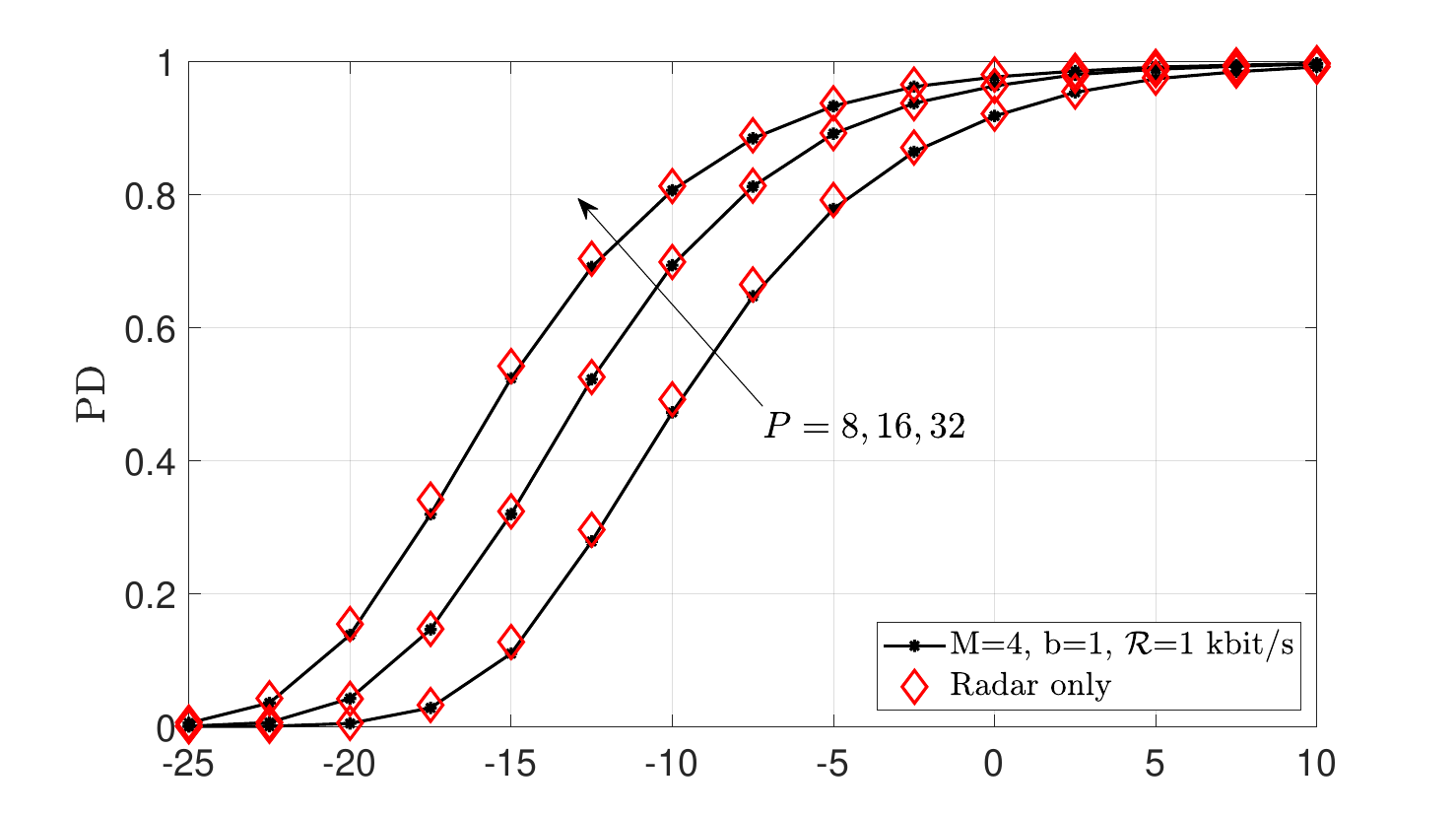}} 
\vspace{-0.2cm}
\centerline{\includegraphics[width=\columnwidth]{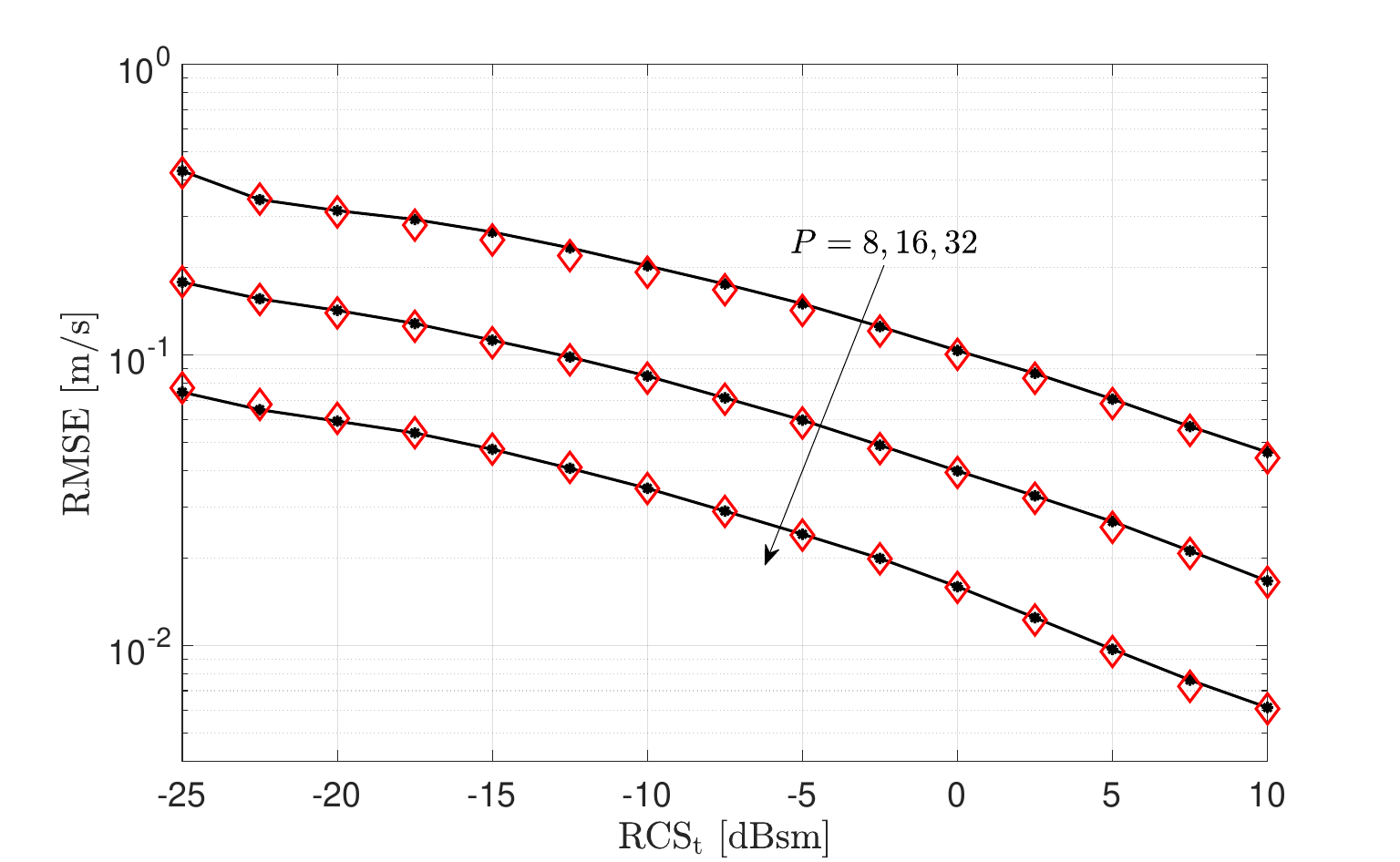}}
\vspace{-0.2cm}
\caption{PD and  RMSE in the estimation of radial velocity  vs ${\rm RCS}_{\rm t}$ with and without the communication function for $P=8,16,32$.}
\vspace{-0.2cm}
\label{fig_PD_RMSE}
\end{figure}

As to the communication function, Fig.~\ref{Fig_bit_error} reports the bit error rate (BER) of each user versus a reference signal-to-noise ratio, defined as
\begin{equation}
\mathrm{SNR}_{\rm u} = \frac{\mathcal{P}\Delta G_{\rm ris}(\bar{\bm{\phi}}_{\rm re,u}) \bigl|\bm{u}_{\rm ris}\transp(\bar{\bm{\phi}}_{\rm re,u})\mathrm{diag}\{\bm{g}\}\bm{s}_{\rm re}\bigr|^2 \sigma^2_{\rm u}  }{L_{\rm re, u} \sigma^2_{\rm com}}, 
\end{equation}
where $\bar{\bm{\phi}}_{\rm re,u}$ is the central direction in $\mathcal{F}_{\rm re,u}$; we consider here various values of $M$ (the number of pulses per time slot) and $b$ (the number of bits per time slot), that together determine the transmission rate $\mathcal{R}=b/(MT)$ in each half-space. It is observed  that different tradeoffs between transmission and error rates can be obtained. For fixed $M$, BER decreases by reducing $b$, at the price of reducing $\mathcal{R}$; for the same $\mathcal{R}$, a larger $M$ should be preferred, as the received signal energy increases. Finally, BER on the reflective side is lower than the one on the other side since the angles of departure of the channel paths fall here closer to the main beam of the STAR-RIS pointing towards the prospective target (see also Fig.~\ref{Fig_01}).

As to the sensing function, the considered performance measures are the probability of declaring the hypothesis $\mathcal{H}_2$ when $\mathcal{H}_2$ is actually true (called the detection probability, or $\mathrm{PD}$) and the root mean square error (RMSE) in the estimation of the target radial velocity.
Fig.~\ref{fig_PD_RMSE} reports $\mathrm{PD}$ and RMSE versus ${\rm RCS}_{\rm t}$ for $P=8,16,32$. We assume here $M=4$ and $b=1$, which provides the largest rate with the shortest time slot in the considered setup; also, $M=4$ is the most favorable choice regarding the coherence time of the user channel (see Sec.~\ref{SEC_rx_signal_user}). The performance when the communication function is not present is also included for comparison. It is observed that adding the communication function has a marginal impact on radar performance. Similar results are obtained with other values of $M$ and $b$ and are omitted due to space constraints.

\section{Conclusions}\label{SEC_Conclusions}
In this study, we have considered an ISAC transceiver, where a STAR-RIS acts as a dual-function transmitter by superimposing space-time modulation on a pulse train emitted by a feeder. The radar receiver relies on a PESA covering both half-spaces and a single digital channel. The spatial response of the STAR-RIS is chosen to illuminate directions that are mirror symmetric. A binary temporal code is employed instead to embed communication messages and distinguish the target echoes from the two sides of the STAR-RIS. The results demonstrate that adding the communication function has a marginal impact on radar performance. We are currently considering the optimization of the STAR-RIS and PESA beampatterns in the presence of clutter.


\end{document}